\documentclass[a4paper,iopams,showpacs,showkey,10pt,final]{iopart}

\usepackage[USenglish]{babel}
\usepackage[utf8]{inputenc}
\usepackage[T1]{fontenc}
\usepackage{amssymb}
\usepackage{epsf}
\usepackage{color}
\usepackage{colordvi}
\usepackage[dvips,final]{graphicx}
\usepackage{textcomp}
\usepackage{cite}

\def\textsubscript#1{\lower .33ex\hbox{\scriptsize #1}}
\def\mLEED{\hbox{\textmu{}-LEED}}

\def\micron{\hbox{\textmu{}m}}
\def\deg{\textdegree}
\def\Celsius{\textcelsius}
\def\mesh#1#2{#1\texttimes#2}
\def\Liii{L\textsubscript{3}}

\def\ie{i.\,e.}

\def\etal{{\itshape et al.\/}}
\def\CoGe#1#2{Co\textsubscript{#1}Ge\textsubscript{#2}}
\def\CoGeii{\CoGe{}{2}}

\def\upperRomannumeral#1{\uppercase\expandafter{\romannumeral#1}}
\def\type#1{type~\upperRomannumeral{#1}}
\def\typed#1{\hbox{type-\upperRomannumeral{#1}}} 

\def\reddish{reddish}

\graphicspath{{./}{../fig_leed/}{../fig_afm/}{../fig_xas/}{../fig_rsm/}}
\begin{document}


\title[Morphology and chemical composition 
       of cobalt germanide islands on Ge(001)]%
      {Morphology and chemical composition 
       of cobalt germanide islands on Ge(001):
       in-situ nanoscale insights into contact 
       formation for Ge-based device technology}

\def\BRE{\textsuperscript1}
\def\FFO{\textsuperscript2}
\def\BTU{\textsuperscript4}
\def\BRC{\textsuperscript3}
\def\FFOandBTU{\textsuperscript{2,4}}
\author{M Ewert\BRE{}, Th Schmidt\BRE{}, J I Flege\BRE{}, I Heidmann\BRE{},
        T Grzela\FFO{},\\ W M Klesse\FFO{}, 
        M Foerster\BRC{}, L Aballe\BRC{},
        T Schroeder\FFOandBTU{}, and J Falta\BRE}
\address{\BRE{}
    Institute of Solid State Physics,
    University of Bremen,
    Otto-Hahn-Allee 1,
    28359 Bremen,
    Germany
}
\address{\FFO{}
    IHP,
    Im Technologiepark 25,
    15236 Frankfurt (Oder),
    Germany
}
\address{\BRC{}
    ALBA Synchrotron Light Facility,
    Carretera BP 1413,
    km 3.3,
    Cerdanyola del Vall\`es,
    Barcelona 08290,
    Spain
}
\address{\BTU{}
    BTU Cottbus-Senftenberg,
    Institute fo Physics and Chemistry,
    Konrad-Zuse-Str.~1,
    03046 Cottbus,
    Germany
}
\ead{tschmidt@ifp.uni-bremen.de}
\vspace*{5pt}%
\begin{indented}
  \item[Date:]\ \rm\raggedright \today
\end{indented}

\begin{abstract}
The reactive growth of cobalt germanide on Ge(001) was
investigated by means of {\itshape in-situ} x-ray
absorption spectroscopy photoemission electron microscopy
(XAS-PEEM), micro-illumination low-energy electron diffraction (\mLEED),
and {\itshape ex-situ} atomic force microscopy (AFM).
At a Co deposition temperature of 670\Celsius{},
a rich morphology with different island shapes and dimensions
is observed, and a correlation between
island morphology and stoichiometry is found.
Combining XAS-PEEM and \mLEED{}, we were able to identify
a large part of the islands to consist of \CoGeii{},
with many of them having an unusual epitaxial relationship: 
\CoGeii$[\bar110](111)$\,$\parallel$\,Ge$[\bar110](001)$.
Side facets with (112) and (113) orientation have been
found for such islands.
However, two additional phases were observed, most likely
\CoGe57 and \CoGe{}{}. The occurrence of these intermediate
phases is promoted by defects, as revealed by
comparing growth on Ge(001) single crystals and
on Ge(001)/Si(001) epilayer substrates.
\end{abstract}
\submitto{\NT}
\maketitle
\ioptwocol


\def\symbolip{
  \def\vvsize{.4em}%
  \def\hhsize{.4em}%
  \def\lwidth{.2pt}%
  \def\frame{%
    \vbox to 0pt{%
      \vss
      \hrule height\lwidth width\hhsize
      \vskip-\lwidth
      \hbox to\hhsize{\vrule width\lwidth height\vvsize\hfill\vrule width\lwidth height\vvsize}%
      \vskip-\lwidth       
      \hrule height\lwidth width\hhsize
    }%
  }
  \def\solid{\hbox to 0pt{\vbox to 0pt{\vss\hrule width\hhsize height\vvsize}\hss}}%
  \kern .3em\raise .1em\hbox{%
    \resizebox{.3em}{.2em}{%
      \rotatebox[origin=c]{45}{%
        \hbox to\hhsize{%
          {\color[rgb]{0.941,0.824,0.667}\solid}
          {\color[rgb]{0.353,0.235,0.078}\frame}
        }
      }
    }
  }
}

\def\symboliiip{
  \hbox{%
    \footnotesize
    \hbox to 0pt{%
      {\color[rgb]{0.823,0.863,0.901}$\bullet$}
      \hss
    }%
    \hbox{\color[rgb]{0.235,0.294,0.314}$\circ$\hss}
  }%
}


\long\def\figi{%
  \begin{figure}[tb]
    \includegraphics[width=\columnwidth]{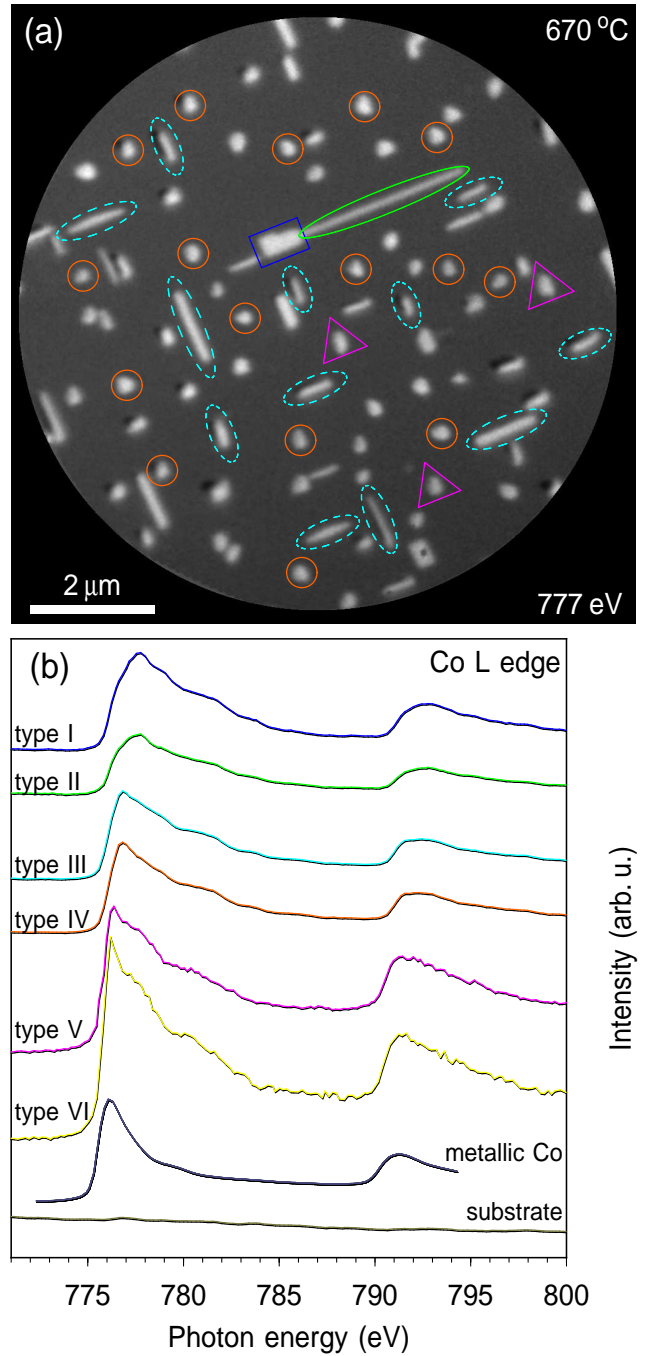}
    \caption{\label{fig:XAS-PEEM}%
     (a) XAS-PEEM image after Co deposition
          on a Ge(001) epilayer, showing islands of various types,
          marked with specific symbols (see text). 
     (b) Local x-ray absorption spectra 
         for \typed{1} to \typed{6} islands, metallic cobalt, and
         for the region between islands (from top to bottom).
         The metallic spectrum has been
         taken from a Co layer deposited at RT (without annealing). 
     }
  \end{figure}
}


\long\def\figii{%
  \begin{figure}[tb]
    \includegraphics[width=\columnwidth]{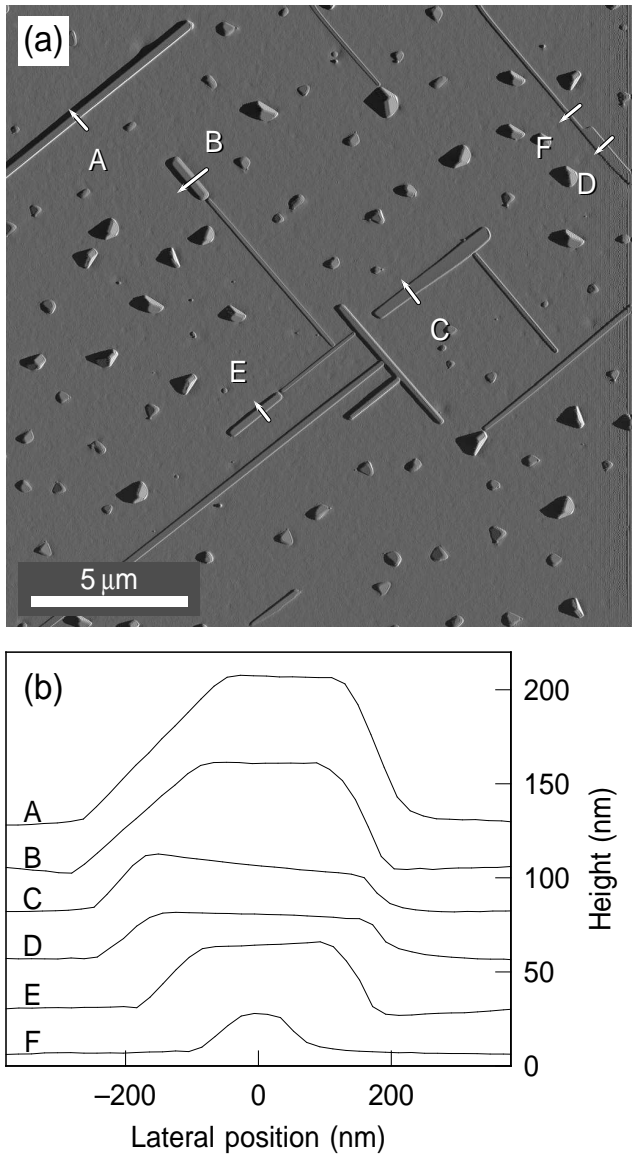}
    \caption{\label{fig:AFM}%
     (a) AFM image after Co deposition 
         on a Ge(001) single crystal and
     (b) several height profiles taken on different germanide islands
         as marked by white arrows and labels A--F in (a).
            Note that islands A--E are of \type{1}, while
            F is a \typed{2} islands. The compact islands 
            on this surface are all of \type{5}.
    }
  \end{figure}
}

\long\def\figiii{%
  \begin{figure*}[tb]
    \includegraphics[width=1.0\textwidth]{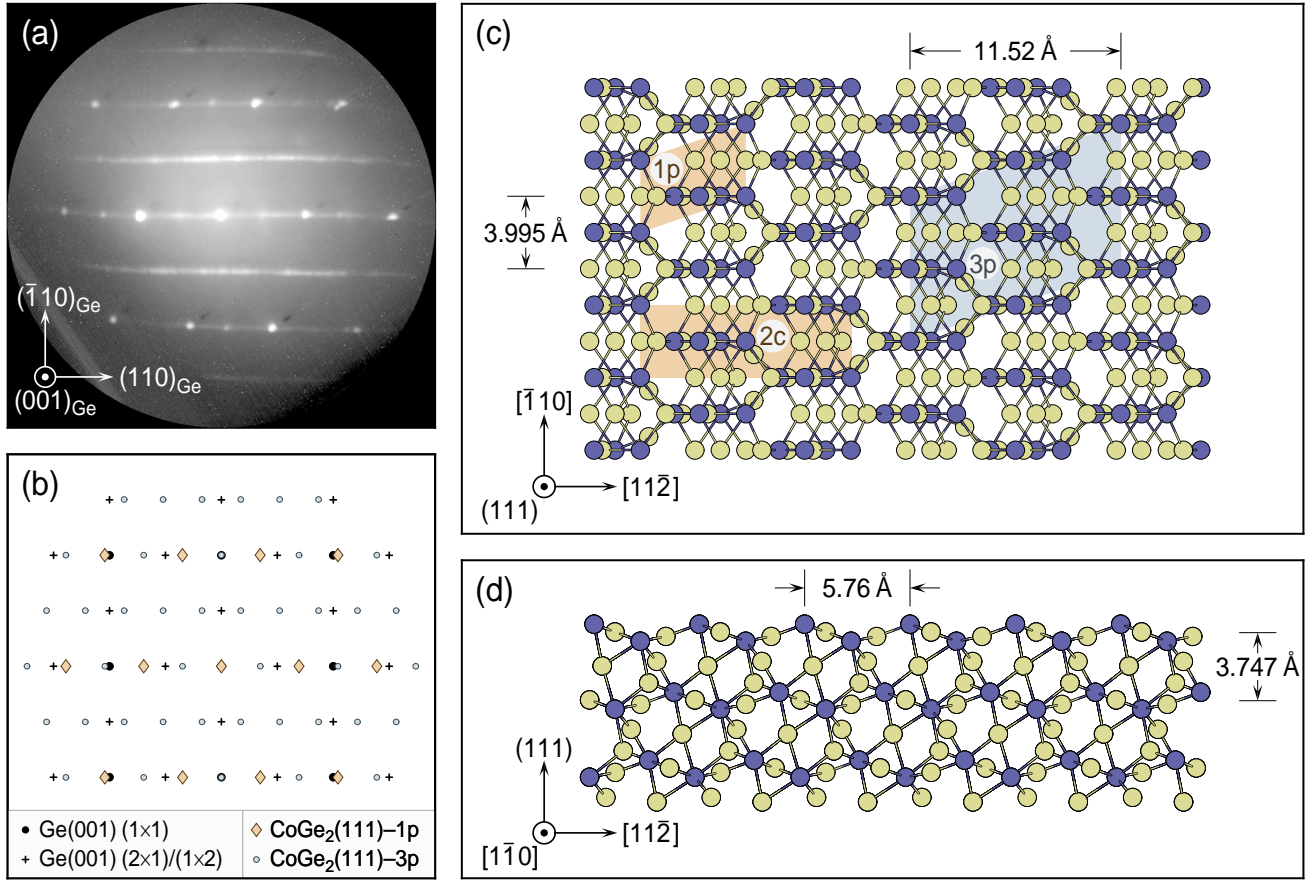}
    \caption{\label{fig:leed_structure}%
     (a) Micro-LEED pattern taken on the edge 
         of a \typed{1} island grown
         on a Ge(001) epilayer,
         averaged over an electron energy range 
         from 3 to 33\,eV.
     (b) LEED spot positions expected for different
         unit meshes: 
         large black dots ($\bullet$) dots 
         correspond to Ge(001)-(\mesh11); 
         plus signs~({\bfseries+}) mark superstructure spots
         from  Ge(001)-(\mesh12) and (\mesh21); 
         large \reddish{}  diamonds~(\protect\symbolip)
         originate from the small oblique unit mesh labeled
          ``1p'' in the top view onto the Co-terminated 
         \CoGeii(111) surface in frame (c); 
         small blue dots~(\protect\symboliiip) 
         mark the additional spots expected for the
         large oblique unit mesh labeled ``3p'' in (c).
     (c) Top view and (d) side view of the 
         bulk-terminated \CoGeii(111)
         surface according to the bulk structure reported in 
         Ref.~\citen{Schubert_NW1948}.
         Blue (dark) spheres correspond to Co, 
         amber (bright) spheres to Ge atoms. 
   }
 \end{figure*}
}


\long\def\figiv{%
  \begin{figure}[tb]
    \includegraphics[width=\columnwidth]{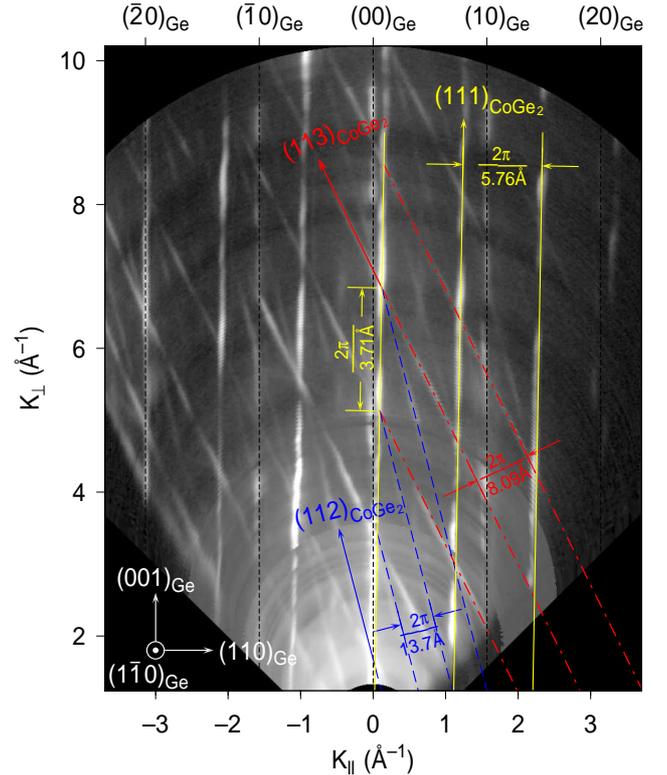}
    \caption{\label{fig:RSM}%
      Reciprocal-space map in the Ge ($1\bar10$)-($001$) plane,
      taken at the edge of a \typed{1} island. 
   }
 \end{figure}
}


\long\def\tabi{
  \begin{table*}[bt]
    \tabcolsep 2\tabcolsep
    \def\arraystretch{1.15}%
    \caption{\label{tab:dimensions}%
      Typical morphologies of different island types as determined from
      LEEM, XAS-PEEM, and AFM.}
      \def\ft{flat-top}
      \def\peaked{peaked}
      \def\epionly{$^{\star}$}
      \def\rare{$^{\star\star}$}
   \vspace{.5ex}%
   \begin{tabular}{lcccc}
      \hline
                       & length (\textmu m) & width (nm) & height (nm) & shape \\
      \hline
      \hline
      \type{1}         & 0.5 -- 5    & 220 -- 700 & 32 -- 65 & elongated, \ft    \\
      \hline
      \type{2}         & 2.7 -- 11.5 & 220 -- 300 & 18 -- 50 & elongated, \peaked \\
      \hline
      \type{3}\epionly & 0.5 -- 1.1  & 230 -- 270 & 10 -- 32 & elongated, \peaked \\
      \hline
      \type{4}\epionly & $\sim$ 0.4  & $\sim$ 320 & 42 -- 65 & compact \\
      \hline
      \type{5}         & $\sim$ 0.4  & $\sim$ 400 & 32 -- 45 & compact, facetted \\
      \hline
      \type{6}\rare    & $\sim$ 1.6  & $\sim$ 180 & $\sim$ 20 & elongated 45\deg{} \\
     \hline
   \end{tabular}\\[2pt]\small
   \hbox to\hsize{\qquad\epionly\ only on epilayer substrates\qquad
                  \rare\   rare occurrence\hfill}
  \end{table*}
}


\long\def\tabii{
  \begin{table*}[tb]
    \tabcolsep 2\tabcolsep
    \def\arraystretch{1.2}%
    \caption{\label{tab:lattices}%
      Bulk unit cell parameters of Ge and some of its cobalt alloys.
    }
    \let\oc=\citen
    \vspace{.5ex}%
    \begin{tabular}{lcccccccc}
      \hline
               & space &  $a$  &  $b$  &  $c$   & $\alpha$ & $\beta$ & $\gamma$ & Ref.\\
               & group & (\AA) & (\AA) & (\AA)  &          &         &          &        \\
      \hline
      Ge       &  227  & 5.658 & 5.658 & 5.658  & 90\deg &  90\deg   & 90\deg{} & \oc{Madelung_1996} \\
      \hline
      \CoGeii  &   41  & 5.65  & 5.65  & 10.80  & 90\deg &  90\deg   & 90\deg{} & \oc{Schubert_NW1948} \\
               &       & 5.670 & 5.670 & 10.796 & 90\deg &  90\deg   & 90\deg{} & \oc{Butts_2004}  \\
      \hline   
      \CoGe57  &  107  & 7.626 & 7.626 & 5.802  & 90\deg &  90\deg   & 90\deg{} & \oc{Audebrand_PD2000}  \\
      \hline
      \CoGe{}{}& 12    & 11.65 & 3.807 & 4.945  & 90\deg & 101.1\deg & 90\deg{} & \oc{Butts_2004} \\
               &       & 11.63 & 3.801 & 4.935  & 90\deg & 100.9\deg & 90\deg{} & \oc{Audebrand_PD2000} \\
      \hline
    \end{tabular}
  \end{table*}
}


\section{Introduction}
In state-of-the-art silicon based complementary
metal oxide semiconductor (CMOS) electronics, the integration
of germanium becomes more and more important, since
it provides materials properties superior to Si.
In addition to an increased performance resulting from a
higher charge carrier mobility\cite{Madelung_1996}, the Ge
band structure can be engineered to exhibit a
quasi-direct bandgap, enabling the fabrication of
laser structures\cite{Claeys_book2007,Liu_OptExpr2007,%
Liu_OptLett2010,Capellini_OptExpr2014,Wirths_NatPhot2015}.
Especially for such high-current applications, one
of the challenges with this materials approach is
to prepare stable, low-resistance contacts that are 
compatible with current processing technologies.
In this respect, a high control over morphology and
chemical composition is required, since inhomogeneous Schottky
barriers limit the overall device performance\cite{Tung_APR2014}.

Based on the experience with nickel and cobalt silicides
as contacts for Si devices, nickel and cobalt germanides
are considered promising for fabrication of high-quality
contacts on Ge. In the following, we address the
growth of cobalt germanides on Ge(001).
In comparison to the Ni--Ge system\cite{Grzela_Nanotechnol2015},
the Co--Ge system exhibits a more complex bulk phase diagram. 
As a consequence, it is important to determine and
to control which phase is formed under which
growth conditions.

The most common scheme to produce 
cobalt germanides on germanium is to
deposit Co which then reacts with the
Ge from the substrate. In general, two
growth techniques are distinguished
in this context:
(i) solid-phase epitaxy (SPE), \ie{},
    Co deposition at room temperature
    and subsequent annealing
and (ii) reactive growth
    by Co deposition at elevated
    temperature.
In the present work we focus on the less-studied
reactive growth. As in the case of cobalt silicide
\cite{Kaenel_PRL1995,Vantomme_APL1999} 
by far not all bulk phases \cite{Ishida_JPhaseEquilibria_1991}
are observed in epitaxial cobalt germanide structures.
Nevertheless, both for reactive growth and
SPE of cobalt germanides, at least three different phases
have been observed \cite{Ashburn_JAP1993,Goldfarb_JVSTB2002}:
\CoGe{}{}, \CoGe57, and \CoGeii.
For Ge-rich conditions, such as in case of small Co deposits
on a Ge substrate, a \CoGeii{} stoichiometry is expected
from the bulk phase diagram \cite{Ishida_JPhaseEquilibria_1991}.
This is beneficial,
since \CoGeii{} has a low ohmic resistance. Moreover, with the
epitaxial relationship \CoGeii[100](001)$\parallel$ Ge[100](001),
growth with very little lattice mismatch (about 0.3\,\%) is
possible. This epitaxial relationship has been
proposed earlier \cite{Goldfarb_JMR2001}, and has experimentally
been observed by Choi \etal{}\ with scanning tunneling microscopy (STM),
who find two different types of surface termination of the germanide,
both with a (\mesh{$\sqrt2$}{$\sqrt2$})R45\deg surface
reconstruction\cite{Choi_JPPC2010}. Despite similar
growth conditions and similar island morphologies,
different surface periodicities have been reported in
other STM studies \cite{Mocking_SS2012,Grzela_JAP2014}.
Such an occurrence of different surface structures may be
related to different, simultaneously occurring epitaxial
relationships. The latter have been demonstrated both
for \CoGe57 and for \CoGeii{}, by transmission electron
microscopy \cite{Sun_APL2005} and,
in particular, by x-ray diffraction (XRD) \cite{DeKeyser_JES2010}.

For SPE, Goldfarb and Briggs propose a phase
sequence\cite{Goldfarb_JVSTB2002}
that does not lead directly to \CoGeii{}. Instead,
at lower annealing temperatures, they propose
the formation of \CoGe57 and possibly still further
intermediate phases  such as \CoGe{2}{} and \CoGe{}{}.
This is in agreement, for instance, with earlier
XRD experiments \cite{Ashburn_JAP1993} by
Ashburn \etal{} 
Under SPE conditions, these authors find a complete
transition from \CoGe57 to \CoGeii{} for rapid annealing at
600\Celsius. In a more recent study \cite{DeKeyser_JES2010},
an even higher transition temperature of 680\Celsius{}
was found.
Contrary, the direct formation of virtually
pure \CoGeii{} has been reported for temperatures as
low as 500\Celsius{} in case of reactive growth 
\cite{Goldfarb_JVSTB2002}.

In recent STM studies \cite{Mocking_SS2012,Grzela_JAP2014},
two different types of cobalt germanide islands
were observed for SPE with
moderate annealing temperatures up to 500\Celsius{}:
(a) islands with a rectangular outline
    and a flat top facet, \ie{}, with a trapezoidal
    cross section, and
(b) elongated, narrow islands with a peaked cross section.
The latter island type has been reported to be
preferred for large island sizes \cite{Mocking_SS2012}
and for higher temperatures \cite{Grzela_JAP2014}.

Despite the fact that the presence of different types
of islands might point to different stoichiometries,
hardly any of the numerous studies on cobalt germanide
growth on Ge(001) provides a link between
island shape and stoichiometry. In the present work
we aim at bridging this gap and focus on the question 
whether island morphology and chemical
composition of cobalt germanide islands on Ge(001) are
correlated, and, in addition,
whether high-temperature growth is a viable approach
to produce single-phase \CoGeii.
For this purpose, we have chosen an experimental 
{\itshape in-situ\/} approach that allows to obtain real-space,
reciprocal-space and spectroscopic information
from the very same structures on a local scale, \ie{}, 
we  used low-energy electron
microscopy (LEEM), micro-illumination
low-energy electron diffraction (\mLEED),
and photoemission electron microscopy
combined with x-ray absorption spectroscopy (XAS-PEEM)
in order to simultaneously investigate the
morphology of the cobalt germanide structures
and probe their chemical composition with a
spatial resolution of a few tens of nanometers.

\section{Experimental}
The XAS-PEEM and \mLEED{} experiments presented in the
following were performed {\itshape in situ} at 
the XPEEM experimental station of the 
CIRCE beamline \cite{Aballe_JSR2015} at the
ALBA Synchrotron Light Facility 
(Spain).

Two types of substrates were used in this study:
On the one hand, single-crystalline
Ge(001) samples cut from Sb-doped wafers were used,
and, on the other hand,  5\,\micron{} thick Ge template layers
epitaxially grown on Si(001), a less perfect substrate
type but technologically more relevant.
The latter were prepared at IHP, Frankfurt (Oder),
Germany, using an approach published
elsewhere\cite{Yamamoto_SSE2011}.
The substrates were cleaned by three cycles of Ar\textsuperscript+ 
ion sputtering
followed by annealing at 800\,\Celsius{} for 30\,min.
The samples were heated by electron bombardment from the
backside, and the temperature was monitored
by a thermocouple attached closely to the sample 
and cross-calibrated by an infrared pyrometer
and, using \mLEED, by the Ge(001) (\mesh21) to (\mesh11) phase
transition temperature \cite{Johnson_PRB1991} at
680\Celsius{}.

Cobalt germanide was grown slightly
below this transition temperature, at about 670\Celsius{}.
For this purpose, cobalt was evaporated at a deposition rate
of 0.34 \AA{} per minute from an electron beam evaporator containing a
Co rod of 99.99\,\% purity, with typical deposits of about
20\,\AA{}.

After growth, the surface was investigated
by LEEM, \mLEED{} \cite{flege_low-energy_2012} 
and XAS-PEEM\cite{Nielson_book2001,Rehr_RMP2000}.
For the latter, the samples were illuminated with
a monochromatic, focused x-ray beam.
The photon energy was scanned across the Co~L{}
absorption edge. Secondary electrons
with a very low kinetic energy (typically 0.5\,eV),
which provide a measure for the local x-ray absorption,
were used for imaging.

{\itshape In-situ} \mLEED{} images were recorded using
a 500\,nm illumination aperture,
in an electron energy range from 2\,eV up to 100\,eV.
Since with LEEM and related techniques it is rather
difficult to assess the height of nanostructures,
complementary {\itshape ex-situ} 
atomic-force microscopy (AFM) investigations of the samples
were conducted at IHP, Frankfurt(Oder), with a
Bruker/Veeco Dimension~5000 instrument operated
under ambient conditions in tapping mode, using an $n^+$-doped
Si cantilever.

\figi
\figii

\section{Results and Discussion}

The XAS-PEEM image in Fig.~\ref{fig:XAS-PEEM}\,(a)
illustrates the surface morphology after Co germanide
growth at 670\Celsius{} on a Ge(001) template layer.
Since the image was recorded with a photon energy slightly
above the Co \Liii{} edge, the contrast reflects the Co
distribution on the surface: Co germanide islands
appear bright, whereas the Ge substrate appears dark.
Obviously, islands have formed with a broad variety 
in size and shape, whereas the remaining
substrate area does not seem to contain any Co, which points
to a Volmer-Weber growth mode.
In Fig.~\ref{fig:XAS-PEEM}, several islands have been marked,
using equal symbols for islands with similar morphology.
From the XAS-PEEM data, together with typical AFM data,
such as shown in Fig.~\ref{fig:AFM}, different types of islands
can be categorized: large and wide rectangular
islands with a flat top facet (\type{1}, blue rectangle in 
Fig.~\ref{fig:XAS-PEEM}); narrow, very long islands
without a flat top facet (\type{2}, green ellipse),
elongated narrow, but rather short and low islands without
flat top facet (\type{3}, dashed cyan ellipses);
very compact and rather high islands (\type{4}, red circles);
and small, irregularly shaped islands (\type{5}, purple triangles).
The anisotropic island types (\type{1} to \type{3}) are
strictly aligned with the $[110]$ or the $[1\bar10]$ 
substrate direction,
\ie{}, they show a 0\deg{} or 90\deg{}
azimuthal orientation. In very few cases, elongated islands
are also found with 45\deg{} orientation (\type{6}, not present
in Fig.~\ref{fig:XAS-PEEM}).
Typical dimensions of the different island types
are summarized in Tab.~\ref{tab:dimensions}.

\tabi

Interestingly, all six island types are observed for
growth on Ge(001) template layer substrates, but only
\typed{1}, \typed{2}, \typed{5}, and \typed{6} islands
are observed when using Ge(001) single-crystal substrates.
Therefore, we conclude that the presence of \typed{3}
and \typed{4} islands on the Ge template layers 
is induced by defects.

Fig.~\ref{fig:AFM} shows typical AFM data obtained from Co germanide
islands grown on a single-crystal substrate.
It can clearly be seen from Fig.~\ref{fig:AFM}\,(a) that the compact
islands have pronounced side facets. However, these \typed{5} islands
exhibit many different azimuthal orientations
in contrast to the elongated islands.

The line profiles in Fig.~\ref{fig:AFM}\,(b) show the presence of 
flat-top (profiles A--E) and peaked (profile F) islands.
The top facets are, in general, not perfectly parallel to the
substrate surface. In Fig.~\ref{fig:AFM}\,(b),
the largest tilt of about 2.0\deg{} occurs for the 
profile of the \typed{1} island labeled ``C''.
The tilt observed here is not attributed to a misalignment
of the instrument; a similar tilt angle has also been observed
with LEED, as described below. It points to a small-angle
grain boundary between Ge and the germanide.

\figiii

Local Co L edge x-ray absorption spectra,
extracted from XAS-PEEM image stacks, reveal the
presence of Co in all island types, as shown in
Fig.~\ref{fig:XAS-PEEM}\,(b).
The regions between the islands, however,
do not show a Co absorption signal (see bottom curve
in Fig.~\ref{fig:XAS-PEEM}\,(b)).
Hence, the regions betwen islands are virtually free of Co,
indicative of a Volmer-Weber growth mode.
Despite the similarity of the absorption
spectra from the different island types,
there are also significant differences.
The spectra for \typed{5} and \typed{6}
islands show a sharp peak very close to the
rising edge. In contrast, the spectra
for \typed{1} and \typed{2} islands
show a broader maximum that appears
at higher photon energy.
In this respect, the spectra
of \typed{3} and \typed{4} islands exhibit
an intermediate shape and peak position.
Hence, three different XAS fingerprints
can be distinguished in Fig.~\ref{fig:XAS-PEEM}\,(b),
pointing to the presence of three different
Co germanide phases.

Before we continue with the interpretation
of this spectroscopic result, we turn to
an independent phase identification by means of
\mLEED{}. For this purpose, a \typed{1} island
has been selected, since this island type is, 
on the hand, one of the most prominent ones,
and, on the other hand, provides a flat top facet,
which simplifies the analysis. The illuminated area
in this experiment contained the center part
of the \typed{1} islands, one of its side facets, and,
for reference, a small part of the substrate region
next to the islands.
Figure~\ref{fig:leed_structure}\,(a) shows the
averaged \mLEED{} pattern obtained over an
energy range from 3 to 33\,eV. 
Due to the constant final kinetic energy in the
imaging column, the LEED spots from a flat surface
do not move on the screen of a LEEM instrument,
in contrast to conventional LEED optics\cite{flege_low-energy_2012}.
Indeed, discrete LEED spots are visible in
Figure~\ref{fig:leed_structure}\,(a), which therefore
can be attributed to the substrate and the top-facet
of the \typed{1} island. This discrete pattern
is superimposed by streaks running from left to right
in Fig.~\ref{fig:leed_structure}\,(a).
These streaks in the energy-averaged image originate
from LEED spots related to the island's side facet(s),
since spots from a tilted surface do move on the screen
with electron energy.
For clarity, we first analyze the discrete part
of the LEED pattern, which is very similar to the sketch
in Fig.~\ref{fig:leed_structure}\,(b).

From the Ge(001) substrate, a square-symmetry LEED pattern
is expected as marked by black dots (integer spots)
and  plus-marks (superstructure spots from 
\mesh21 reconstruction in two 90\deg{} rotational domains)
in Fig.~\ref{fig:leed_structure}\,(b).
Since the illuminated substrate area in the
\mLEED{} experiment was very small, these spots are rather
faint in the experimental LEED pattern shown in
Fig.~\ref{fig:leed_structure}\,(a).
The most intense spots correspond to the
\reddish{} diamonds in Fig.~\ref{fig:leed_structure}\,(b)
and have to be attributed to the top facet of the germanide island.
They form an oblique pattern,
typical of a centered rectangular unit mesh.
Their spacing along the the bottom-up axis of the
LEED pattern (Ge$(\bar110)$ direction) is the
same as for the substrate spots, \ie{}, within the experimental
resolution the germanide growth is pseudomorphic
or lattice-matched with respect to the Ge substrate along 
its $[\bar110]$ direction.
This is also the direction along which the germanide island is
elongated; hence, a low lattice mismatch is energetically
favorable in this direction. 
The best lattice match for low-index planes
and directions for \CoGe{}{} and \CoGe57 that comply with a
rectangular LEED-pattern is obtained for (cf.~Tab.~\ref{tab:lattices})
\CoGe{}{}$[010]$\,$\parallel$\,Ge$[\bar110]$, and
\CoGe57$[100]$\,$\parallel$\,Ge$[\bar110]$,
with a lattice mismatch of almost 5\,\%
along the Ge$[\bar110]$ direction in both cases.
It seems very unlikely that such a large mismatch persists
over such mesoscopic islands.
In contrast, a lattice mismatch of less than 0.3\,\% occurs
for  \CoGeii$[\bar110]$ along Ge$[\bar110]$
(cf.~Tab.~\ref{tab:lattices}).
\tabii
In fact, the complete LEED pattern can be explained in terms
of the \CoGeii{} structure, as described in the
following. We emphasize at this point
that the previously proposed epitaxial
relationship \cite{Goldfarb_JVSTB2002,Choi_JPPC2010} with
\CoGeii(001) parallel to Ge(001) can be excluded in this case,
since this would require a LEED pattern with square symmetry.
Instead, our data reveal the previously unreported
\CoGeii{}$[\bar110](111)$\,$\parallel$\,Ge$[\bar110](001)$
orientation shown in Fig.~\ref{fig:leed_structure}\,(c)
and (d). Figure~\ref{fig:leed_structure}\,(c) is a top view
of the \CoGeii(111) surface, using the
atomic positions within the bulk unit cell as
determined by Schubert and Pfisterer \cite{Schubert_NW1948}.
The primitive surface unit mesh is the large blue 
parallelogram marked ``3p'' in Fig.~\ref{fig:leed_structure}\,(c).
When only the two uppermost atomic layers are considered,
however, the primitive surface unit mesh is given by the
small parallelogram marked ``1p''; alternatively,
the non-primitive, centered rectangular unit mesh
marked ``2c'' can be used. Even when more than two
layers are considered, only few atoms violate the
surface periodicity defined by ``1p'' (or, 
equivalently, ``2c'').
Therfore, the most intense spots of the LEED pattern
will correspond to this reciprocal unit cell 
(marked by diamonds in Fig.~\ref{fig:leed_structure}\,(b)).
The small deviations, leading to the actually larger
``3c'' unit mesh, produce additional spots which are 
expected to be relatively weak 
(marked by blue dots in  Fig.~\ref{fig:leed_structure}\,(b)).
This model agrees very well with the experimental
LEED pattern in Fig.~\ref{fig:leed_structure}\,(a).

\figiv

Figure~\ref{fig:RSM} shows a reciprocal-space map
in the Ge$(001)$-$(1\bar10)$ plane, extracted
from a series of \mLEED{} images taken as a function of
electron energy \cite{MzH_RSI2005,Schmidt_NJP2007}.
Reciprocal-lattice rods with a periodicity of 2$\pi$/4.00\,\AA{}
run along the vertical direction, \ie{}, along the Ge(001)
direction. They are identified as integer reciprocal-lattice
rods of the Ge(001) substrate and marked by dashed
black lines. The half-order rods generated by the
\mesh21 surface reconstruction are hardly visibly.
(Best to be seen is the $(\bar{\frac12} 0)$ rod.)
The most intense rods (solid yellow lines)  are attributed
to the top facet of the \typed{1} island. These rods
are not perfectly aligned along the vertical direction,
but have a small tilt angle of about 1.0\deg{}, similar
to what we observed
with AFM, as mentioned above.
The spacing of these reciprocal-lattice rods
corresponds to a real-space periodicity of
5.76\,\AA, which perfectly matches the periodicity
of the \CoGeii{}(111) surface along its $[11\bar2]$
direction, as depicted in  Fig.~\ref{fig:leed_structure}\,(d).
Faint half-order top-facet rods can also be seen.
Despite the fact that the unit mesh is actually
twice as large (11.52\,\AA)
in the $[11\bar2]$ direction, reflections
corresponding to this periodicity (i.\,e.,
such half-order rods) are forbidden
in the plane of Fig.~\ref{fig:RSM}
due to glide plane symmetry. 
(Note that the projection of the bulk-terminated
real-space lattice into this plane has
truly a periodicity of 5.76\,\AA, as visible
in Fig.~\ref{fig:leed_structure}\,(d).)
Hence, these faint spots might
indicate a surface reconstruction or
originate from multiple scattering.

The orientation of the side facets can be determined
from the inclination angles of the related
reciprocal-lattice rods. Two sets of such rods are
particularly pronounced, with angles (relative to
the \CoGeii(111) rods) of 27.0\deg{} (red dot-dashed lines in
Fig.~\ref{fig:RSM}) and 15.5\deg{} (blue dashed lines).
The respective facets' real-space periodicities
as determined from the rods' spacing are 8.09\,\AA{} and
13.7\,\AA{}.
These results can be explained by the presence of
\CoGeii{} (113) and (112) facets,
for which inclination angles and periodicities of
27.7\deg{} and 16.2\deg, 8.06\,\AA{} and 13.4\,\AA{},
respectively, are expected, in very good agreement
with experiment.
The inclination angles determined with \mLEED{} are 
also in good agreement with the AFM data; for
the \typed{1} islands labeled ``C'' and ``D'' in
Fig.~\ref{fig:AFM}, side facet angles between
16\deg{} and 24\deg{} are found. 
In case of crystallographic facets, the reciprocal facet rods
intersect at bulk Bragg peak conditions \cite{Horn_ZKrist1999}.
Hence, from the vertical distance of these intersections,
the layer spacing of the germanide islands can be
determined. The experimentally obtained value of
3.71\,\AA{} gives further support for 
\typed{1} islands being of (111) orientation, since
in this case an almost identical value for the 
layer spacing of 3.747\,\AA{} is 
expected (see Fig.~\ref{fig:leed_structure}\,(d)).

Having identified \typed{1} islands as \CoGeii{}, 
and turning back to the XAS-PEEM results in
Fig.~\ref{fig:XAS-PEEM}, we can also identify
\typed{2} islands as \CoGeii{}, since they have
virtually identical XAS fingerprints. The
different morphology is probably due to a different
epitaxial relationship. The latter could not be resolved
in our experiments, since \typed{2} islands are very narrow
and, thus, produce only weak \mLEED{} intensity.

Regarding the peak shift of the \typed{1} to the \typed{6}
fingerprints towards the metallic spectrum in Fig.~\ref{fig:XAS-PEEM}\,(b),
it seems reasonable to relate this peak shift to the stoichiometry
of the islands, \ie{}, to assign an increasing peak energy
to increasing Ge content,
since the \CoGeii{} stoichiometry of \typed{1} and \typed{2} islands
is the one with the highest Ge content (and, of course, pure Co has the
lowest Ge content). Following this reasoning, \typed{3} and \typed{4}
islands are likely to consist of \CoGe57, and \typed{5} and
\typed{6} islands of \CoGe{}{}.
Since islands of \type{3} and \type{4} are not observed on
single-crystal Ge(001) substrates but only on epilayers, it is concluded
that the nucleation and growth of
islands with \CoGe57{} stoichiometry is related to defects.

\section{Conclusion}

We have shown that high-temperature reactive growth of
co\-balt germanide on Ge(001) leads to a variety of
island morphologies, including elongated and compact
structures. Moreover, a correlation of island morphology
and cobalt germanide stoichiometry has been revealed
by local XAS. From a detailed \mLEED{} analysis,
one of the most prominent island types was identified
as \CoGeii{}(111), an orientation
that has not been reported so far. These islands
exhibit (112) and (113) side facets.
In addition to \CoGeii{},
different germanide phases have been detected and
are attributed to \CoGe57 and \CoGe{}{}.
Therefore, high-temperature reactive growth
does not seem to be a superior approach to enforce
single-phase cobalt germanide growth.
Remarkably, the occurrence of such intermediate phases
is strongly suppressed on single-crystal substrates and
can thus be attributed to defects.
It should be noted that this is not a general
trend, since on amorphous Ge, for instance,
the formation of \CoGeii{} has been reported without any
\CoGe57 \cite{Opsomer_APL2007}.

With respect to applications, an
important conclusion is that the defect structure and
defect density of Ge layers on Si play a significant role
here and have to be taken into account
when trying to transfer results from single-crystal based
research, i.\,e.\ from the vast majority of
previous literature, into device development.

\section*{References}
\bibliography{coge_v6}

\providecommand{\newblock}{}
\begin{thebibliography}{10}
\expandafter\ifx\csname url\endcsname\relax
  \def\url#1{{\tt #1}}\fi
\expandafter\ifx\csname urlprefix\endcsname\relax\def\urlprefix{URL }\fi
\providecommand{\eprint}[2][]{\url{#2}}

\bibitem{Madelung_1996}
Madelung O (ed) 1996 {\em Semiconductors --- basic data\/}
  {2\textsuperscript{nd}} ed (Berlin: Springer Verlag)

\bibitem{Claeys_book2007}
Claeys C and Simoen E 2007 {\em Germanium-Based Technologies: From Materials to
  Devices\/} 1st ed (Amsterdam: Elsevier Science)

\bibitem{Liu_OptExpr2007}
Liu J, Sun X, Pan D, Wang X, Kimerling L~C, Koch T~L and Michel J 2007 {\em
  Opt. Express\/} {\bf 15} 11272--11277

\bibitem{Liu_OptLett2010}
Liu J, Sun X, Camacho-Aguilera R, Kimerling L~C and Michel J 2010 {\em Opt.
  Lett.\/} {\bf 35} 679--681

\bibitem{Capellini_OptExpr2014}
Capellini G, Reich C, Guha S, Yamamoto Y, Lisker M, Virgilio M, Ghrib A, Kurdi
  M~E, Boucaud P, Tillack B and Schroeder T 2014 {\em Opt. Express\/} {\bf 22}
  399--410

\bibitem{Wirths_NatPhot2015}
Wirths S, Geiger R, von~den Driesch N, Mussler G, Stoica T, Mantl S, Ikonic Z,
  Luysberg M, Chiussi S, Hartmann J~M, Sigg H, Faist J, Buca D and
  Gr{\"u}tzmacher D 2015 {\em Nat. Photonics\/} {\bf 9} 88--92

\bibitem{Tung_APR2014}
Tung R~T 2014 {\em Appl. Phys. Rev.\/} {\bf 1} 011304

\bibitem{Grzela_Nanotechnol2015}
Grzela T, Capellini G, Koczorowski W, Schubert M~A, Czajka R, Curson N~J,
  Heidmann I, Schmidt T, Falta J and Schroeder T 2015 {\em Nanotechnol.\/} {\bf
  26} 385701

\bibitem{Kaenel_PRL1995}
von K\"anel H, Schwarz C, Goncalves-Conto S, M\"uller E, Miglio L, Tavazza F
  and Malegori G 1995 {\em Phys. Rev. Lett.\/} {\bf 74}(7) 1163--1166

\bibitem{Vantomme_APL1999}
Vantomme A, Degroote S, Dekoster J, Langouche G and Pretorius R 1999 {\em Appl.
  Phys. Lett.\/} {\bf 74} 3137--3139

\bibitem{Ishida_JPhaseEquilibria_1991}
Ishida K and Nishizawa T 1991 {\em J. Phase Equilibria\/} {\bf 12} 77--83

\bibitem{Ashburn_JAP1993}
Ashburn S~P, {\"O}zt{\"u}rk M~C, Harris G and Maher D~M 1993 {\em J. Appl.
  Phys.\/} {\bf 74} 4455--4460

\bibitem{Goldfarb_JVSTB2002}
Goldfarb I and Briggs G~A~D 2002 {\em J. Vac. Sci. Technol. B\/} {\bf 20}
  1419--1426

\bibitem{Goldfarb_JMR2001}
Goldfarb I and Briggs G~A~D 2001 {\em J. Mater. Res.\/} {\bf 16} 744--752

\bibitem{Choi_JPPC2010}
Choi J, Lim D~K, Kim Y and Kim S 2010 {\em J. Phys. Chem. C\/} {\bf 114}
  8992--8996

\bibitem{Mocking_SS2012}
Mocking T~F, Hlawacek G and Zandvliet H~J~W 2012 {\em Surf. Sci.\/} {\bf 606}
  924--927 ISSN 00396028

\bibitem{Grzela_JAP2014}
Grzela T, Koczorowski W, Capellini G, Czajka R, Radny M~W, Curson N, Schofield
  S~R, Schubert M~A and Schroeder T 2014 {\em J. Appl. Phys.\/} {\bf 115}
  074307

\bibitem{Sun_APL2005}
Sun H~P, Chen Y~B, Pan X~Q, Chi D~Z, Nath R and Foo Y~L 2005 {\em Appl. Phys.
  Lett.\/} {\bf 86} 071904

\bibitem{DeKeyser_JES2010}
De~Keyser K, Van~Meirhaeghe R~L, Detavernier C, Jordan-Sweet J and Lavoie C
  2010 {\em J. Electrochem. Soc.\/} {\bf 157} H395--H404

\bibitem{Aballe_JSR2015}
Aballe L, Foerster M, Pellegrin E, Nicolas J and Ferrer S 2015 {\em J.
  Synchrotron Rad.\/} {\bf 22} 745--752

\bibitem{Yamamoto_SSE2011}
Yamamoto Y, Zaumseil P, Arguirov T, Kittler M and Tillack B 2011 {\em
  Sol.-State Electron.\/} {\bf 60} 2--6

\bibitem{Johnson_PRB1991}
Johnson A~D, Norris C, Frenken J~W~M, Derbyshire H~S, MacDonald J~E,
  Van~Silfhout R~G and Van Der~Veen J~F 1991 {\em Phys. Rev. B\/} {\bf 44}
  1134--1138

\bibitem{flege_low-energy_2012}
Flege J~I, Tang W~X and Altman M~S 2012 {\em Characterization of Materials\/}
  ed Kaufmann E~N (Hoboken: John Wiley \& Sons) chap Low-Energy Electron
  Microscopy 2nd ed

\bibitem{Nielson_book2001}
Als-Nielson J and McMorrow D 2001 {\em Elements of Modern X-ray Analysis\/}
  (Hoboken: John Wiley \& Sons)

\bibitem{Rehr_RMP2000}
Rehr J~J and Albers R~C 2000 {\em Rev. Mod. Phys.\/} {\bf 72} 621--654

\bibitem{Schubert_NW1948}
Schubert K and Pfisterer H 1948 {\em Die Naturwissenschaften\/} {\bf 35} 222
  ISSN 0028-1042, 1432-1904

\bibitem{Butts_2004}
Butts D~A and Gale W~F 2004 {\em Smithells metals reference book\/} ed Gale W~F
  and Totemeir T~C (Oxford: Elsevier Butterworth-Heinemann) chap Crystal
  chemistry 8th ed

\bibitem{Audebrand_PD2000}
Audebrand N, Ellner M and Mittemeijer E~J 2000 {\em Powder Diffraction\/} {\bf
  15}(2) 120--122 ISSN 1945-7413

\bibitem{MzH_RSI2005}
Meyer~zu Heringdorf F~J and Horn-von Hoegen M 2005 {\em Rev. Sci. Instrum.\/}
  {\bf 76} 085102

\bibitem{Schmidt_NJP2007}
Schmidt T, Clausen T, Flege J~I, Gangopadhyay S, Locatelli A, Mentes T~O, Guo
  F~Z, Heun S and Falta J 2007 {\em New J. Phys.\/} {\bf 9} 392

\bibitem{Horn_ZKrist1999}
{Horn{-}von{ }Hoegen} M 1999 {\em Z. Krist.\/} {\bf 214} 591--628 and 684--721

\bibitem{Opsomer_APL2007}
Opsomer K, Deduytsche D, Detavernier C, Van~Meirhaeghe R~L, Lauwers A, Maex K
  and Lavoie C 2007 {\em Appl. Phys. Lett.\/} {\bf 90} 031906

\end{thebibliography}
\bibliographystyle{iopart-num}
\end{document}